\documentclass{llncs}
\usepackage{graphicx}
\usepackage{pb-diagram}
         \usepackage{floatrow}
\usepackage{setspace}
\begin{document}

\title{Modelling Requirements for Content Recommendation Systems}
\author{Sarah Bouraga \inst{1,2} \and Ivan Jureta  \inst{1,2,3} \and St\'ephane Faulkner  \inst{1,2} }
\institute{Department of Business Administration, University of Namur \and PReCISE Research Center, University of Namur \and Fonds de la Recherche Scientifique -- FNRS, Brussels \\ \email{$\{$sarah.bouraga, ivan.jureta, stephane.faulkner$\}$@unamur.be}}

\maketitle

\begin{abstract}
This paper addresses the modelling of requirements for a content Recommendation System (RS) for Online Social Networks (OSNs).  On OSNs, a user switches roles constantly between content generator and content receiver.  The goals and softgoals are different when the user is generating a post, as opposed as replying to a post.  In other words, the user is generating instances of different entities, depending on the role she has: a generator generates instances of a ``post", while the receiver generates instances of a ``reply".  Therefore, we believe that when addressing Requirements Engineering (RE) for RS, it is necessary to distinguish these roles clearly.  

We aim to model an essential dynamic on OSN, namely that when a user creates (posts) content, other users can ignore that content, or themselves start generating new content in reply, or react to the initial posting.  This dynamic is key to designing OSNs, because it influences how active users are, and how attractive the OSN is for existing, and to new users.  We apply a well-known Goal Oriented RE (GORE) technique, namely i-star, and show that this language fails to capture this dynamic, and thus cannot be used alone to model the problem domain.  Hence, in order to represent this dynamic, its relationships to other OSNs' requirements, and to capture all relevant information, we suggest using another modelling language, namely Petri Nets, on top of i-star for the modelling of the problem domain.  We use Petri Nets because it is a tool that is used to simulate the dynamic and concurrent activities of a system and can be used by both practitioners and theoreticians \cite{murata1989petri}.  
\end{abstract}

\section{Introduction} \label{sec:Chap_IStar_Extension_introduction}

Over the years, new types of systems have been developed, raising new challenges for Requirements Engineering (RE). Abstractions necessary for modelling of the corresponding new requirements may have to be updated.  However, there is little research of OSN from the perspective of RE.

One of these new system types are Online Social Networks (OSNs). OSNs refer to \cite{ellison2007social} as: 
\begin{quote}
 ``Web-based services that allow individuals to (1) construct a public or semi-public profile within a bounded system, (2) articulate a list of other users with whom they share a connection, and (3) view and traverse their list of connections and those made by others within the system.  The nature and nomenclature of these connections may vary from site to site."  
\end{quote}

OSNs differ from traditional ISs because the behavior of one user has an impact on the behavior of other users and of the system itself. More specifically, both OSNs and users generate content. When a user shares what we call an event type (an event type is an activity generated by a user that can produce a notification to this user's friends, for instance ``Share a photo"), the user's friends have a choice: they can decide to reply to that event type or not. This decision has an impact on the information that is exchanged on the system. We can also observe that the amount and the order in which the event types are notified to the users vary depending on the OSNs. For instance, Tumblr, Twitter, and Instagram notify each user with all the event types generated by the user's friends. Others, such as Facebook, seem to apply some sort of a filtering, or also called recommendation technique in order to decide which event types are notified, and in what order the event types are displayed to the user. This dynamic and these specificities make the OSNs a particular class of systems; the requirements of which might not be modeled using existing requirements modelling languages.

More specifically, on OSNs, a user switches roles constantly between content generator and content receiver.  The goals and softgoals are different when the user is generating a post, as opposed as replying to a post.  In other words, the user is generating instances of different entities, depending on the role she has: a generator generates instances of a ``post", while the receiver generates instances of a ``reply".  Therefore, we believe that a RS, which needs to do content recommendation, needs to see these roles as separate.  
 
In this paper, we address the modelling of requirements for a Recommendation System (RS) for OSN.  More specifically, we are trying to model the problem domain we introduced above, that is of content recommendation on OSNs.  GORE is a popular problem-oriented RE approach \cite{quartel2009goal}, and we choose a well-known GORE technique, namely i-star (i*).  We choose i-star over KAOS, another well-known GORE technique, because we believe that i* is more appropriate when - as in our case - individuals' intentions, actions and roles are central to analysis.  We want to represent this problem in one single model, because how the observed dynamic is designed into the OSN influences if the OSN will satisfy some of the basic user goals (such as, enjoy the OSN), and will influence the extent to which important softgoals are satisficed.  We want to have one model where we can both do the analysis of the recommendation mechanism, and represent how it influences user goals and softgoals. 

This question leads to another question: If we cannot use one single requirements language to model the dynamics, then what new concepts and/or relations do we need to use together with the original language?  We show that i* cannot capture all relevant information; thus we are using two modelling languages together in order to represent the requirements.  More specifically, the aims and corresponding contributions of this paper are threefold: (i) we identify the shortcomings of i-star for the modelling of the problem domain, that is for the content recommendation on OSNs particular purpose of doing requirements engineering of OSNs, (ii) we propose to use another modelling language on top of i-star to address these identified shortcomings, namely Petri-Nets, for the modelling of the dynamics observed on OSNs, and (iii) we illustrate this contribution with requirements models.
 
In order to approach these questions, we apply the following methodology. Firstly, we model the problem domain using the existing concepts proposed by i-star. Secondly, we analyze the shortcomings of the language for the modelling of content recommendation on OSNs.  Thirdly, we propose to use the Petri Nets on top of i-star so that the identified shortcomings can be addressed, and we will propose new models including this other language.

The remainder of the paper is organized as follows. Section \ref{sec:Chap_IStar_Extension_motivating_ex} introduces the motivating example.  In Section \ref{sec:Chap_IStar_Extension_overview}, we give an overview of our approach, and in Section \ref{sec:Chap_IStar_Extension_model_languages} we describe each language that we plan to use.  We apply the specified languages to our example, and discuss the results in Sections \ref{sec:Chap_IStar_Extension_application} and \ref{sec:Chap_IStar_Extension_discussion} respectively.  Section \ref{sec:Chap_IStar_Extension_rel_work} reviews the related work.  Finally, Section \ref{sec:Chap_IStar_Extension_ccl} concludes the paper and discusses the future work. 

\section{Motivating Example} \label{sec:Chap_IStar_Extension_motivating_ex}
We consider two OSN users, User A and User B. User A and User B are ``friends" on the OSN. User A shares something on the OSN, that is, she generates an event type. The OSN has to decide if the event type should be notified to User B. If the event type is notified to User B, then User B has to decide whether to reply to the event type. For instance, if User A shares a photo on the OSN, and if the photo is notified to User B, then the latter has to decide whether she will like, or comment the photo for instance.
 
\begin{figure}[htbp]
\begin{center}
 \includegraphics[width=3in]{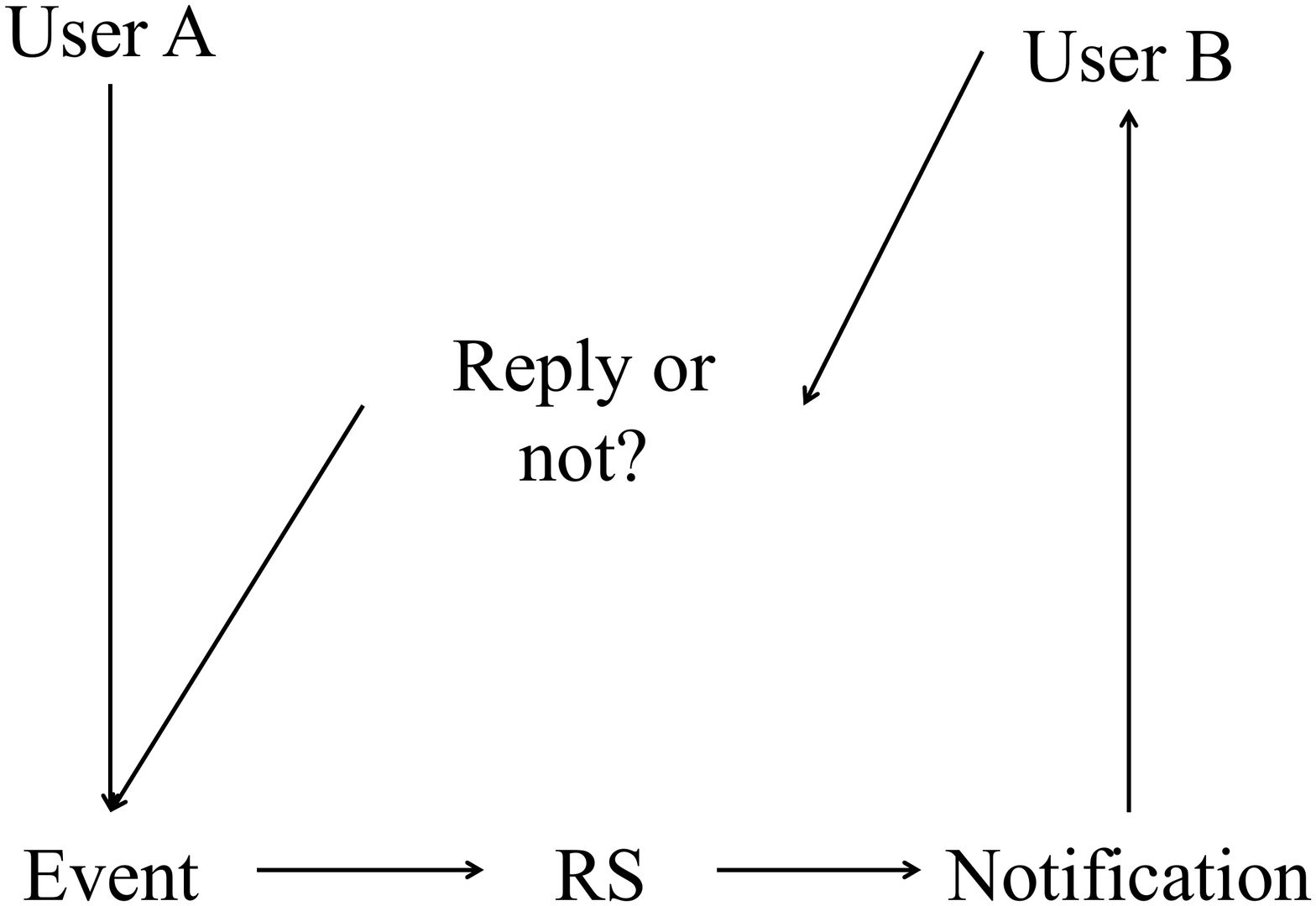}
\caption{First Round of Dynamics}
\label{fig:Chap_IStar_Extension_cycle}
\end{center}
\end{figure}

If User B decides to reply to the event type, then her reply amounts to an event type, and she now acts as generator, that is, if she replies, then User B has generated an event to which other users may choose to reply.  Hence, the mechanism goes on. The OSN decides if the reply/event type should be notified to User A. If the event type is notified, then User A has to decide if she wants to reply to the event type generated by User B. And the mechanism continues.  The first round of this dynamic is represented in Figure \ref{fig:Chap_IStar_Extension_cycle}.

\section{Overview} \label{sec:Chap_IStar_Extension_overview}

The research question is: How can we represent the requirements for RS in one single i-star diagram?  The reason we want it in one i-star model, is that how this dynamic is designed into the OSN influences if the OSN will satisfy some of the basic user goals (such as, enjoy the OSN), and will influence the extent to which important softgoals are satisficed.  We want to have one model where we can both do the analysis of the recommendation mechanism, and represent how it influences user goals and softgoals. This question leads to another question: What new concepts and/or relations do we need to use together with those of i-star to show the dynamics represented in Figure \ref{fig:Chap_IStar_Extension_cycle}?
 
When trying to model the motivating example using only the original concepts of i-star, we are confronted to a limitation of i-star.  We can model the fact that a user can take on both roles, the role of the generator and the role of the receiver.  However, we cannot model the ``role swap" that occurs, nor the element that triggers the swap.  Put another way, we cannot model the notion of mechanism/dynamic, the first round of which is represented in Figure \ref{fig:Chap_IStar_Extension_cycle}.
 
We did not want to introduce a new concept to the i-star language, or any other language.  Indeed, we did not want to add a new extension to the many that already exist in the literature.  We believe that we can model our problem domain with existing concepts.  And in doing so, our results would be more convenient because it will not require the acquisition of new knowledge by the readers.  So instead, we turn to a layer mechanism.  We propose to use a hierarchical model in order to introduce this notion of dynamics to the model.  More specifically, we have a base layer, and a second layer where the concept of dynamics is modeled. 
 
An efficient way to introduce the notion of dynamics is to use Petri Nets.  Hence, we use the latter to address the problem we are confronted with i-star.  Thus, the second layer consists of a Petri Nets model. 
 
The connection between layers happen if the User B (the Receiver) decides to reply to the Event Type.  If this trigger does not occur, then the base layer is sufficient.  Otherwise, if it does occur, the second layer is activated. 
 
Our approach is not to extend i-star with new concepts, that is, we are not changing the meta-model of i-star.  However, we aim to represent the dynamic in one layer, have the i-star model in the other layer, and then have relationships which connect the two layers.

\section{The Selected Modelling Languages} \label{sec:Chap_IStar_Extension_model_languages}

\subsection{I-Star Layer} \label{sec:Chap_IStar_Extension_istar_layer}

\subsubsection{Why I-Star?} 

First of all, the reason why we choose the GORE approach is that it is a popular approach for problem-oriented RE, that facilitates reasoning about the purpose of a proposed solution \cite{quartel2009goal}.  We are in a problem-oriented context since we are trying to model and analyze the problem domain, namely content recommendation on OSNs; and we aim to consider reasoning on these models.  Therefore, the GORE approach is appropriate. 
 
Secondly, i-star is a popular GORE technique.  Table \ref{tab:Chap_IStar_Extension_reasoning} summarizes various well-known GORE techniques for our purpose here.

\begin{table}
\begin{center}
 \caption{Reasoning}
 \label{tab:Chap_IStar_Extension_reasoning}
  \begin{tabular}{l|l}
 \hline
\textbf{Languages} &  \textbf{Motivation} \\
\hline
\textbf{I-Star} \cite{yu1997towards} & Modeling and reasoning about organizations.  The SD model \\
& focuses on the intentional relationships between actors.  The SR \\
& model focuses on the intentional constructs within each actor.\\
\textbf{KAOS} \cite{dardenne1993goal} & Used for requirements acquisition.  Allows the capture of the  \\
& ``what" requirements, ``why", ``who", and ``when" aspects. \\
\textbf{NFR} \cite{chung2009non} & Modelling and using of non-functional requirements during the \\
& development process.  Proposes the notion of satisficing. \\
\textbf{GBRAM} \cite{anton1998use} & Process for the discovery, analysis, and refinement of goals.\\
\hline
  \end{tabular}
\end{center}
\end{table}

We choose i-star over KAOS because we believe that the former is more expressive than the latter.  We do not choose NFR because we are not only interested in softgoals.  Finally, we prefer to have a visual syntax, thus GBRAM is not appropriate.  The fact that i-star is one of the most used language was also a factor in our decision \cite{horkoff2015using,decreus2009practical}.

\subsubsection{Description of the I-Star Language} 

The motivation behind the i* framework is the modeling and reasoning about organizations: their environments, as well as their ISs. Two modeling elements compose the framework, namely the Strategic Dependency (SD) model and the Strategic Rationale (SR) model. They are used, respectively, for the description of the dependency relationships between various actors; and for the description of stakeholder interests and concerns, and how they might be dealt with by different configurations of systems and their environments \cite{yu1997towards}.  

One of the model composing the i* framework is the Strategic Dependency (SD) model, focusing on the intentional relationships between actors, which allows a deeper understanding of the whys. Various dependency types exist, differentiating between the types of freedom and constraint \cite{yu1997towards}.  

The Strategic Rationale (SR) model provides a way to model the intentional constructs within each actor. The intentional elements (goals, tasks, resources, and softgoals) in the SR model are linked by (i) means-ends, and (ii) task-decomposition relationships. The former explain ``why an actor engage in some tasks, pursue a goal, need a resource, or want a softgoal"; while the latter describe hierarchically the intentional elements making up a routine \cite{yu1997towards}.  

\subsubsection{Application of the I-Star Language} 

User A plays the \textit{Role of} Generator, and the \textit{Role of}  Receiver.  User B plays the \textit{Role of} Generator, and the \textit{Role of}  Receiver.  

We will detail the intentional constructs of the OSN, where ET stands for Event Type, and ETI stands for Event Type Instance:

\begin{itemize}
 \item The OSN has the internal task of creating and maintaining the desire/interest for the user to use the OSN; decomposed into: 
	 \begin{itemize}
		\item The goal ``Propose relevant recommendations", 
		\item The goal ``Allow users to generate ETs/Propose features", 
		\item The softgoal ``Mitigate information overload", 
		\item The softgoal ``Encourage dynamics", 
		\item The softgoal ``Generate more content", 
		\item The softgoal ``Improve user experience", 
	\end{itemize}
\item The  goal ``Allow users to generate ETs/Propose features" can be achieved by: 
	  \begin{itemize}
		\item Allowing users to post profile ETs, 
		\item Allowing users to post link ETs, 
		\item Allowing users to post content ETs, 
		\item Allowing users to post recommendations ETs, 
		\item Allowing users to post privacy ETs, 
		\item Allowing users to post connection ETs, 
	 \end{itemize}

\item The  goal ``Propose relevant recommendations" can be decomposed into: 
	  \begin{itemize}
		\item The task ``Gather ETI", achieved by: 
	  	 \begin{itemize}
			\item The task ``Gather original/starting ETI", 
		 		 \begin{itemize}
				\item Depends on the Generator for the resource ``Original ETI", 
				 \end{itemize}
			\item The task ``Gather reply to an original/starting ETI", 
					 \begin{itemize}
					\item Depends on the Generator for the resource ``Reply ETI", 
					\end{itemize}
		\end{itemize}
	\item The task ``Gather User information", achieved by: 
		 \begin{itemize}
		\item ``Analyze relations between user", 
		\item ``Analyze the content of ETI"
		 \end{itemize}
	\end{itemize} 

\item The OSN depends on the RS to propose relevant recommendations.  The RS has the internal task of ``Choose user to receive notifications of user's ETI", decomposed into: 
	  \begin{itemize}
		\item The task ``Get user information"
, achieved by: 
			\begin{itemize}
			\item Depends on the OSN for the resource ``User information"
			 \end{itemize}	
		\item The task ``Apply decision heuristic", achieved by: 
			\begin{itemize}
			\item The task ``Decide to notify the user with the ET"
			\item The task ``Decide to not notify the user with the ET"
			\item Achieved by the task ``Decision heuristic based on empirical research on users' preferences for ETs"
			 \end{itemize}	
		\end{itemize}

\item The softgoal ``Generate more content" is impcted by: 
	  \begin{itemize}
		\item Allow users to generate more ETIs (+)
		\item Gather original ETI (+)
		\item Gather reply ETI (+)
		\item Decide to notify (+)
		\item Decide to not notify (-)
		\end{itemize}

\item The softgoal ``Refine user profile" (refine the information about user's preferences for some event types.  For instance, if the user often replies to a specific event type (let's say, a photo), then the OSN can add this information to the user profile.  The next time one of the user's friends shares a photo, the OSN can decide to notify the user more easily.)  is impacted by: 
	  \begin{itemize}
		\item Gather user information (+)
		\item Gather reply ETI (+)
		\end{itemize} 

\item The softgoal ``Encourage dynamics" (encourage exchanges between users) is impcted by: 
	  \begin{itemize}
		\item Choose users to receive notification of user's ETI (-)
		\item Propose features (+)
		\item Gather reply ETI (+)
		\end{itemize} 

\item The softgoal ``Improve user experience" is impcted by: 
	  \begin{itemize}
		\item Propose features (+)
		\item Propose relevant recommendations (+)
		\end{itemize} 

\item The softgoal ``Mitigate information overload" is impcted by: 
	  \begin{itemize}
		\item Propose features (-)
		\item Gather original ETI (-)
		\item Gather reply ETI (-)
		\item Decide to notify (-)
		\item Decide to not notify (+)
		\end{itemize} 
\end{itemize}

We will now turn to the intentional constructs of the other actors:

\begin{itemize}
 \item The Generator has the internal task of ``Generating ET", achieved by:
	 \begin{itemize}
		\item The task ``Generate original ETI", 
		\item The task ``Generate reply ETI", 
	\end{itemize}
\item The Generator depends on the OSN for the resource ``Feature".
 \item The Receiver has the internal task of ``Use the OSN passively", decomposed into:
	 \begin{itemize}
		\item The task ``Receive notifications of user's ETI", 
			 \begin{itemize}
			\item Depends on the RS for the resource ``Notification"
			\end{itemize}
		\item The task ``Evaluate ETI", achieved by: 
			\begin{itemize}
			\item ``Evaluate ETI as relevant", 
				\begin{itemize}
				\item Means to the end ``React to user's ETI"
				\item The softgoal ``Notifications are relevant to user's friends who receive them" contributes positively to the task
				\end{itemize}
			\item ``Evaluate ETI as irrelevant", 
				\begin{itemize}
				\item Means to the end ``Do not react to user's ETI"
				\end{itemize}
			\end{itemize}
	\end{itemize}

\item If the Receiver decides to reply to an ETI, then the Receiver depends on the Generator, that is, on the other role, to act on this decision and to generate the reply ETI.  
\end{itemize}

\begin{figure}[htbp]
\begin{center}
 \includegraphics[width=3in]{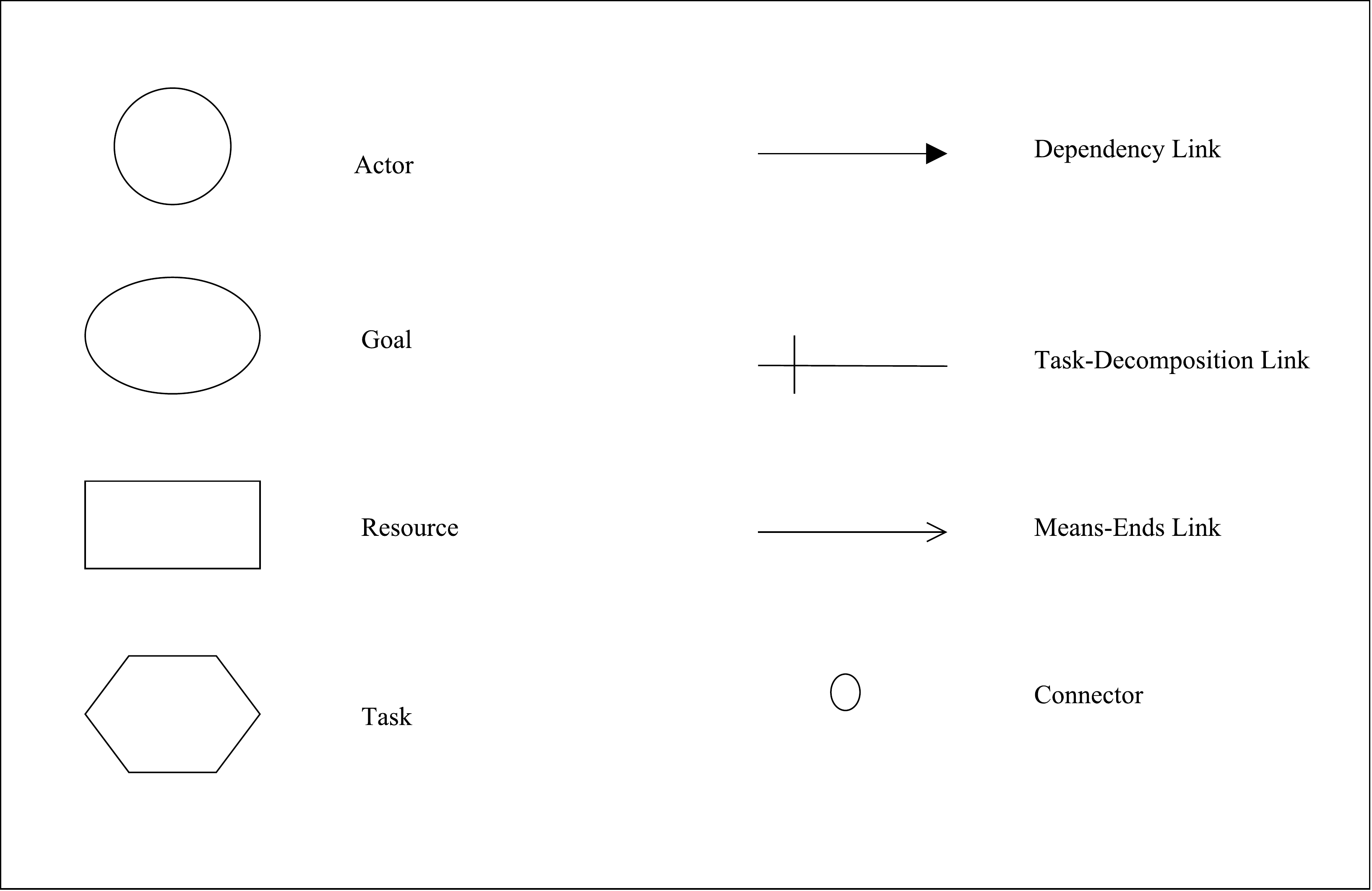}
\caption{Legend}
\label{fig:Chap_IStar_Extension_legend}
\end{center}
\end{figure}

\begin{figure}[htbp]
\begin{center}
 \includegraphics[width=5in]{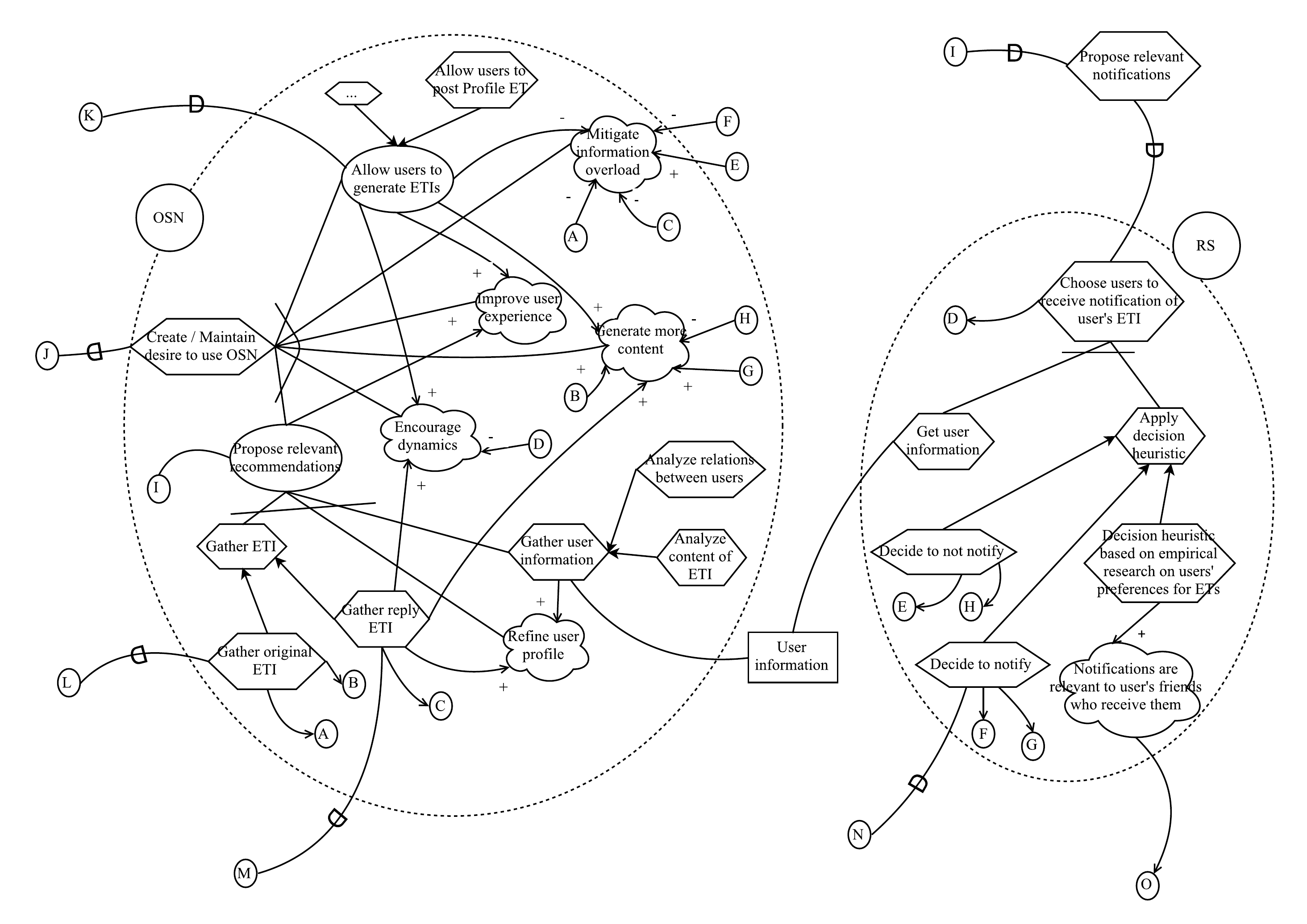}
\caption{I-Star Layer: Strategic Rationale Model - Part 1}
\label{fig:Chap_IStar_Extension_SR1}
\end{center}
\end{figure}

\begin{figure}[htbp]
\begin{center}
 \includegraphics[width=5in]{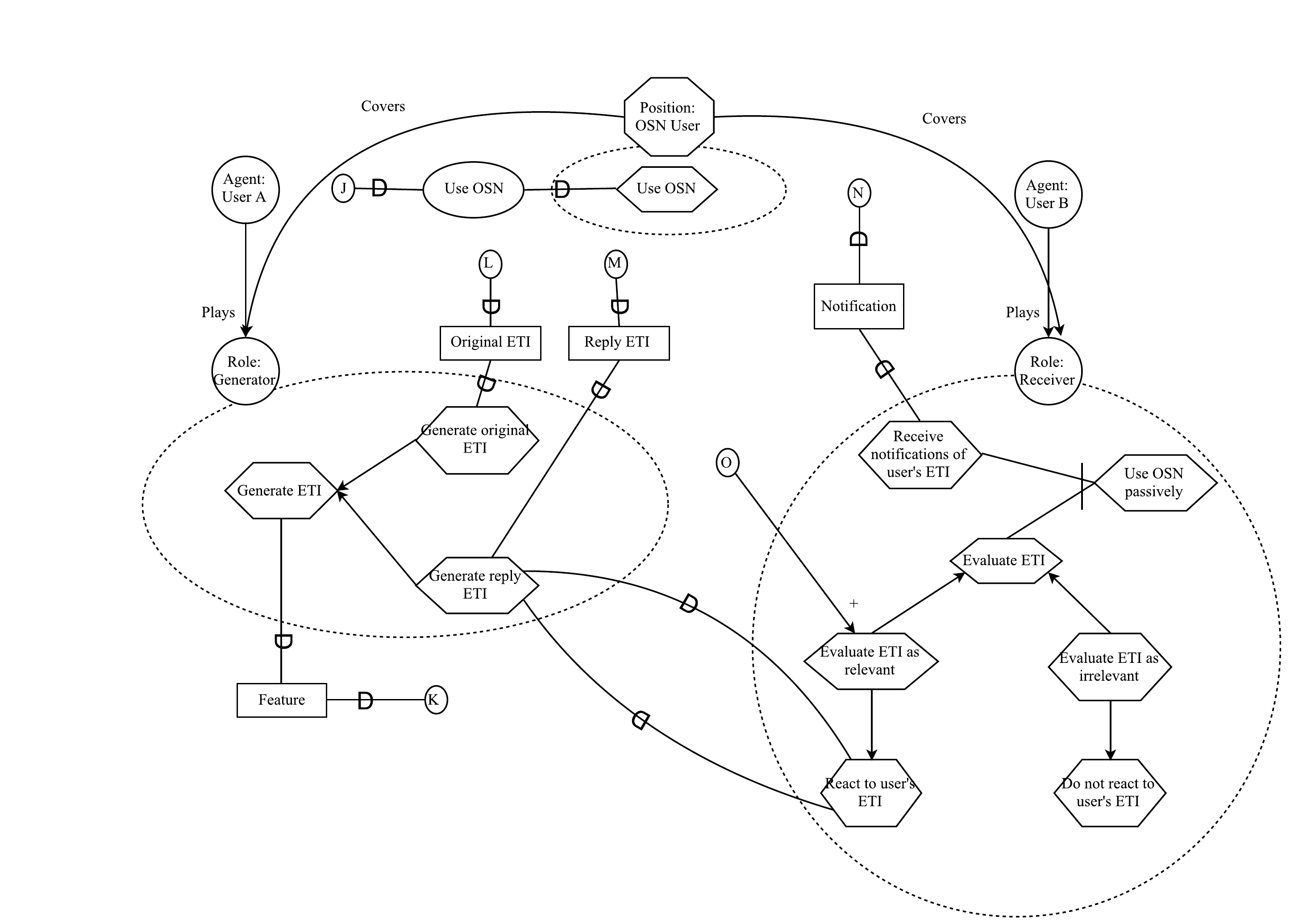}
\caption{I-Star Layer: Strategic Rationale Model - Part 1}
\label{fig:Chap_IStar_Extension_SR2}
\end{center}
\end{figure}

\subsection{Petri Net Layer} \label{sec:Chap_IStar_Extension_pn_layer}

\subsubsection{Why Petri Net?} 

Many theories allow to model the notion of dynamics: sequence diagrams, activity diagrams, state machine, etc.  We choose to use Petri Nets because it is a simple way to model dynamics.  Petri Nets offer a simple, and formal method for dynamics representation.  The other theories either lack a sound or formal semantics, and/or propose a more complicated meta-model.  For our purpose here, we do not need a lot of expressivity; we only need the possibility to model the flow we can observe. 

\subsubsection{Description of the Petri Net} 

A Petri net is a particular kind of directed graph, together with an initial state called the initial marking, M$_0$ \cite{murata1989petri}.  
 
Petri net consists of two kinds of nodes: (i) places, and (ii) transitions.  Arcs go from a place to a transition; or they go from a transition to a place. Graphically, places are represented as circles, and transitions as bars or boxes.  Each place is assigned a non negative integer by a particular marking (state).  Graphically, \textit{k} black dots (tokens) are represented in place \textit{p}. A marking is designated by \textit{M},  an m-vector, where \textit{m} is the total number of places. The p${th}$ component of \textit{M}, indicated by \textit{M(p)} is the number of tokens in place \textit{p}.  

In modeling, using the concept of conditions and events, places are used to model conditions, and transitions to model events. A transition has a certain number of input and output places.  The former represent the preconditions of the event, while the latter represent its postconditions.  If a token is present in a place, then the condition represented by the place is considered as \textit{true}.

Murata argues that: 
\begin{quote}
``In order to simulate the dynamic behavior of a system, a marking in a Petri net is changed according to the following transition (firing) rule: 
\begin{enumerate}
\item A transition t is said to be enabled if each input place p of t is marked with at least w(p,t) tokens, where w(p,t) is the weight of the arc from p to t.
\item An enabled transition may or may not fire (depending on whether or not the event actually takes place).
\item A firing of an enabled transition t removes w(p,t) tokens from each input place p of t, and adds w(t,p) tokens to each output place p of t, where w(t,p) is the weight of the arc from t to p."
\end{enumerate}
\end{quote}

\subsubsection{Application of the Petri Net} 

In our case study, the places represent the nodes of the first layer of the SR Model:
\begin{itemize}
\item p$_1$: Gather original ET
\item p$_2$: Analyze ET
\item p$_3$: Choose to not notify
\item p$_4$: Choose to notify
\item p$_5$: Decide to not react
\item p$_6$: Decide to react
\item p$_7$: Agent User A becomes Receiver
\item p$_8$: Agent User B becomes Generator
\item p$_9$: Generate reply
\item p$_{10}$: Reply
\item p$_{11}$: Gather reply
\item p$_{12}$: Generate more content
\item p$_{13}$: Encourage dynamics
\item p$_{14}$: Refine user profile
\item p$_{15}$: Mitigate information overload
\end{itemize}

\begin{figure}[htbp]
\begin{center}
 \includegraphics[width=5in]{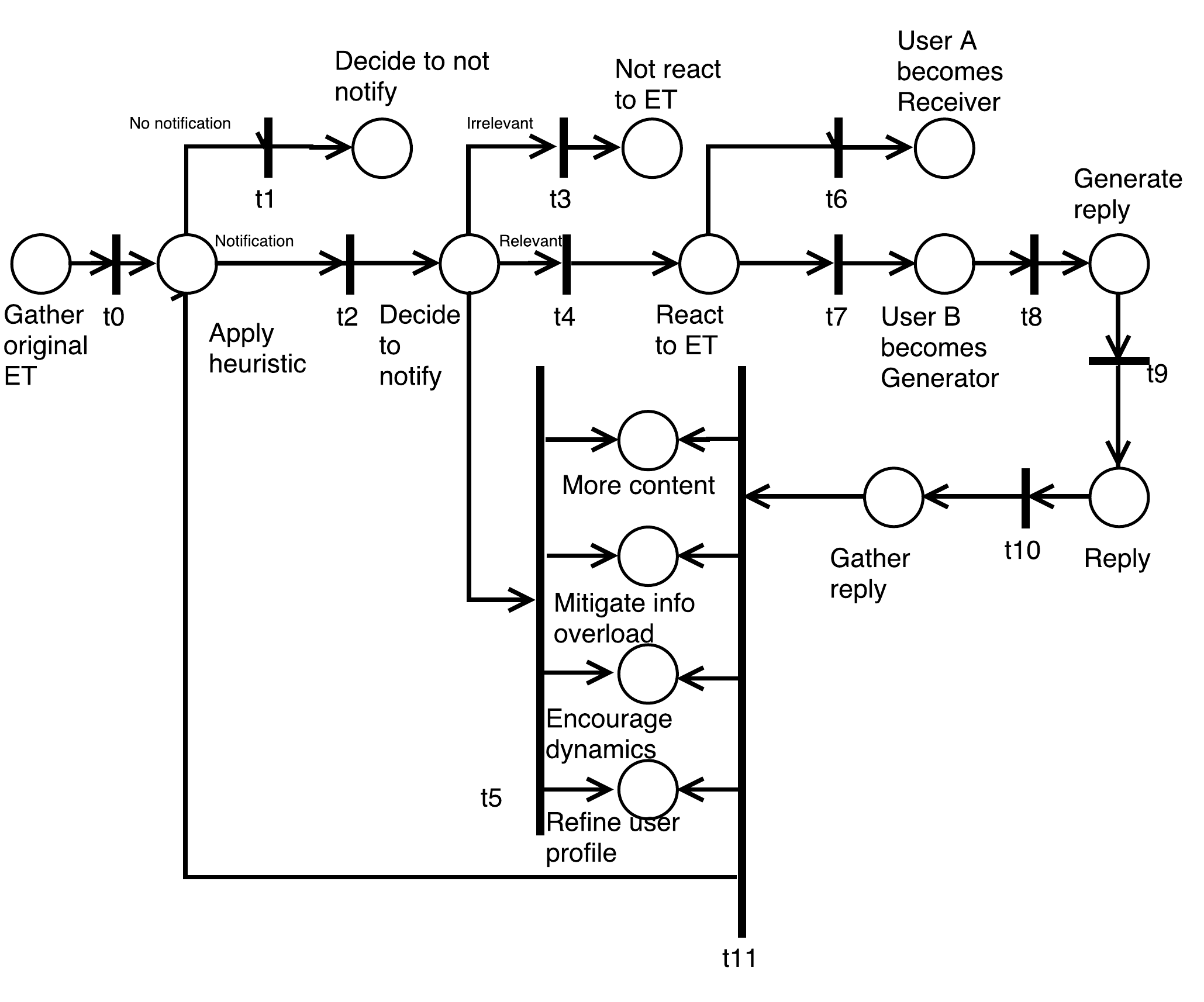}
\caption{Petri Net Layer}
\label{fig:Chap_IStar_Extension_PN}
\end{center}
\end{figure}

The initial marking M$_0$ is:
\begin{displaymath}
M_0 = < 1, 0, 0, 0, 0, 0, 0, 0, 0, 0, 0, 0, 0, 0, 0 >
\end{displaymath}

\begin{figure}[htbp]
\begin{center}
 \includegraphics[width=5in]{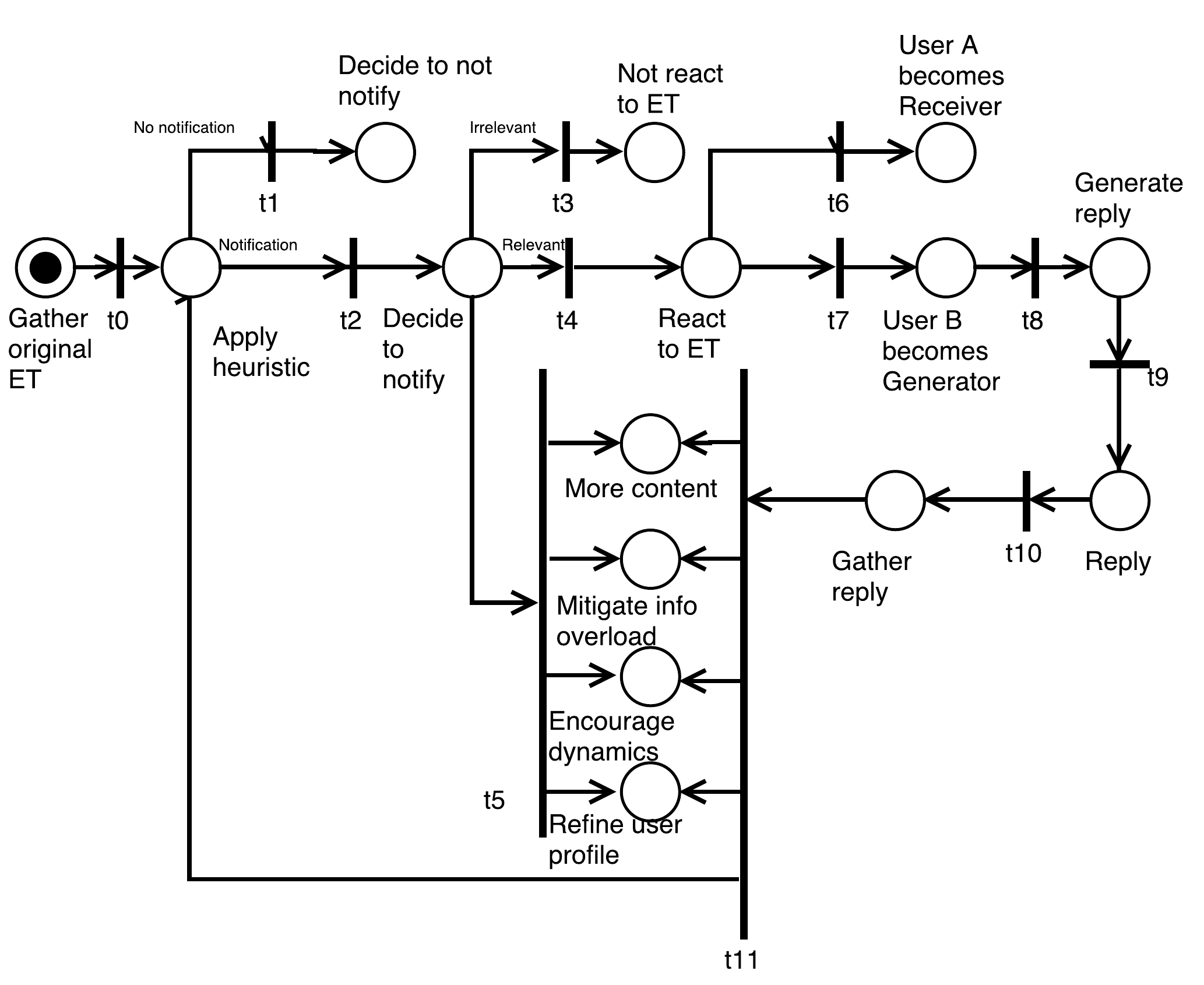}
\caption{Petri Net Layer with Initial Marking}
\label{fig:Chap_IStar_Extension_PN}
\end{center}
\end{figure}

The first transition, t$_0$, is enabled: there is a token on its input arc.  When the transition fires, it consumes the token from p$_1$, and it adds a token to each of its output place.  This results in M$_1$.
 
In our example, once the OSN gathers the original ET, the RS applies the decision heuristic.  If the RS decides to notify the receiver with the ET, then it adds a token to each of its output place p$_4$.  This results in M$_1$ and M$_2$: 
 
\begin{displaymath}
M_1 = < 0, 1, 0, 0, 0, 0, 0, 0, 0, 0, 0, 0, 0, 0, 0 >
\end{displaymath}

\begin{displaymath}
M_2 = < 0, 0, 0, 1, 0, 0, 0, 0, 0, 0, 0, 0, 0, 0, 0 >
\end{displaymath}

If the receiver considers the ET as relevant and wants to react to it, then we have M$_3$:

\begin{displaymath}
M_3 = < 0, 0, 0, 0, 0, 1, 0, 0, 0, 0, 0, 0, 0, 0, 0 >
\end{displaymath}
 
In our example, once the receiver decides to react to the ET, we want to model a change in roles.  User A becomes Receiver, and User B becomes Generator.
 
\begin{displaymath}
M_4 = < 0, 0, 0, 0, 0, 0, 1, 1, 0, 0, 0, 0, 0, 0, 0 >
\end{displaymath}

The eighth transition, t$_8$, is enabled: there is a token on its input arc.  When the transition fires, it consumes the token from p$_8$, and it adds a token to its output place, that is p$_9$. This results in M$_5$.
 
In our example, the task ``Generates a reply" occurs.
 
\begin{displaymath}
M_5 = < 0, 0, 0, 0, 0, 0, 0, 0, 1, 0, 0, 0, 0, 0, 0 >
\end{displaymath}

The ninth transition, t$_9$, is enabled: there is a token on its input arc.  When the transition fires, it consumes the token from p$_9$, and it adds a token to its output place, that is p$_{10}$. This results in M$_6$.
 
In our example, the resource ``Reply" is generated.
 
\begin{displaymath}
M_6 = < 0, 0, 0, 0, 0, 0, 0, 0, 0, 1, 0, 0, 0, 0, 0 >
\end{displaymath}

The tenth transition, t$_{10}$, is enabled: there is a token on its input arc.  When the transition fires, it consumes the token from p$_{10}$, and it adds a token to its output place, that is p$_{11}$. This results in M$_7$.
 
In our example, the task ``Gather reply" occurs.
 
\begin{displaymath}
M_7 = < 0, 0, 0, 0, 0, 0, 0, 0, 0, 0, 1, 0, 0, 0, 0 >
\end{displaymath}

The eleventh transition, t$_{11}$, is enabled: there is a token on its input arc.  When the transition fires, it consumes the token from p$_{11}$, and it adds a token to each of its output places, that is p$_{12}$, p$_{13}$, p$_{14}$, and p$_{15}$. It also adds a token to p$_2$.  This results in M$_8$.
 
In our example, once the task ``Gather reply" has occurred, the following softgoals are contributed to: Generate more content, Encourage dynamics, Refine user profile, and Mitigate information overload.  Also, the RS has to apply the decision heuristic to the reply. 
 
\begin{displaymath}
M_8 = < 0, 1, 0, 0, 0, 0, 0, 0, 0, 0, 0, 1, 1, 1, 1 >
\end{displaymath}

\subsection{Connection Between Layers} \label{sec:Chap_IStar_Extension_connection}
How do these layers connect?  The base layer, that is the i-star layer, represents the various elements that can occur in an OSN.  The second layer, that is the Petri Net layer, represents the dynamic found in the content recommendation context of an OSN.  The second layer is activated when a specific trigger occurs.  This trigger is the generation of an original event type by a user.  Afterwards, if the RS decided to notify the ET, the receiver has to decide on the relevance, and on her desire to reply to the event type. 
 
Graphically, how does the connection between both layers occur?  Once the trigger occurs, the symbol of the base layer lift up to the second layer.  More specifically, we replace the circles of the Petri Nets with the corresponding symbol of the base layer, that is with the i-star concepts.  Hence, the model reads more easily; because we directly see to what symbol the circles of the Petri Net correspond. 
 
Nevertheless, we do not insert new symbols or new concepts.  All the symbols and concepts are known and belong to the i-star or Petri Net languages.  We just use the Petri Net formalism to sequence the i-star symbols.

\section{Application to the Case Study} \label{sec:Chap_IStar_Extension_application}

We will use the i-star model as the base layer.  On the second layer, we will use the Petri Nets model.  On the first layer, we manipulate only the original concepts of i-star.  On the second layer, we manipulate the concepts of both languages: i-star and the Petri Nets.  More specifically, for the second layer, we keep the notions of transitions and places, but we replace the symbol of ``places" with the corresponding symbol of i-star.  Thus, the changes are the following:
\begin{itemize}
\item The places representing a task are modeled by an hexagon
\item The places representing an agent are modeled by a circle
\item The places representing a resource are modeled by a rectangle
\item The places representing a softgoal are modeled by a cloud
\end{itemize}
 
\begin{figure}[htbp]
\begin{center}
 \includegraphics[width=5in]{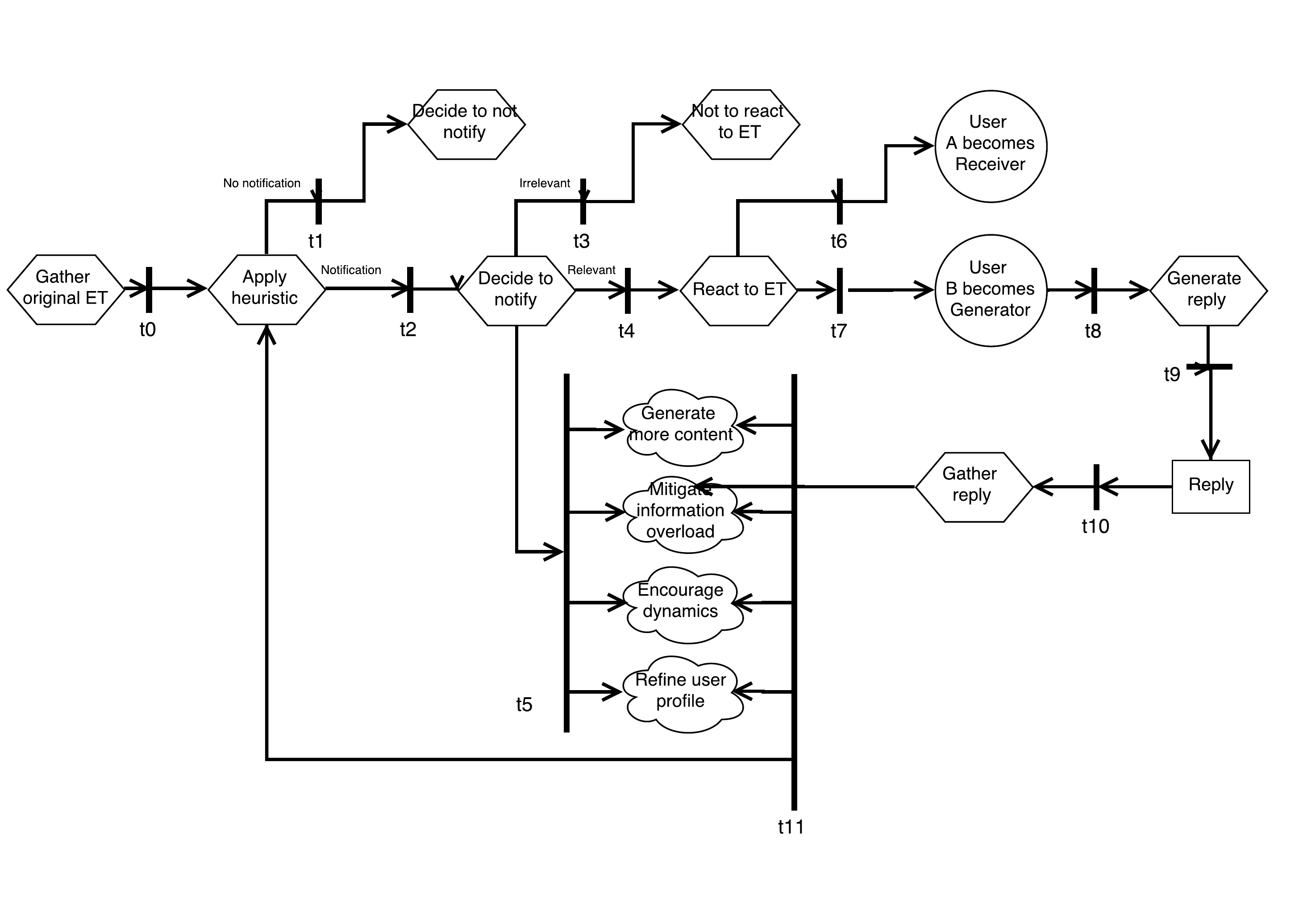}
\caption{Connection Between Layers}
\label{fig:Chap_IStar_Extension_PN_IStar}
\end{center}
\end{figure}

In our application, the circle p$_1$, which represents the task ``React to user's ET", is replaced by an hexagon.  This is also the case for p$_4$ and p$_6$.  The places p$_2$ and p$_3$ are represented by circles.  The place p$_5$ is modeled using a rectangle.  Finally, the places p$_7$ to p$_{10}$ are modeled by clouds.  The ``new" Petri Net model is represented in Figure \ref{fig:Chap_IStar_Extension_PN_IStar}.  

The first layer is thus composed of Figures \ref{fig:Chap_IStar_Extension_SR1} and \ref{fig:Chap_IStar_Extension_SR2}; while the second layer consists of Figure \ref{fig:Chap_IStar_Extension_PN_IStar}. 
 
This swap allows a clear connection between the two layers.  The relevant content for the modelling of the dynamics is elevated from the i-star layer to the second layer; where we use the Petri Net procedure to model the observed dynamics. 
 
If the Receiver evaluates the ETI as relevant, it causes to react to this ETI.   The latter leads to: (i) a change in the role, that is, User B becomes the Generator, and User A becomes the Receiver; and (ii) the generation of a reply ETI by the Generator.  Once the Generator has generated a reply, it leads to a new resource ``Reply ETI".  The OSN will gather this reply, and it will have an influence on various softgoals.  More specifically, the reply will contribute positively to the following softgoals: Encourage dynamics, Refine user profile, and Generate more content.  However, the reply will contribute negatively to the softgoal Mitigate information overload. 
 
In our example, we can see that the Observer depends on the Generator to act on her desire to reply; and the Generator depends on the Observer to decide to reply in order to generate the reply.  The Generator has the internal task of generating a reply; and the Observer has the internal task of deciding to reply.  However, while the roles change, it is the same User who has the internal tasks of deciding to reply, and of actually generating the reply.
 
In our model, we show the dependencies in the base layer; and the actual switch in roles is shown in the second layer.  If the receiver decides to react to the generator's ET, then in the second layer, a token is present in the corresponding place ``React to the user's ET".  The token travels through the net, and the markings evolve as the token moves along. 
 
We also see that the event has implications on the softgoal.  A reply contributes positively to the softgoals ``Generate more content", ``Encourage dynamics", and ``Refine user profile", and it contributes negatively to ``Mitigate information overload".  We see a reinforcing loop for the softgoals the reply contributes positively to.
 
We will consider the softgoal ``Generate more content".  The User A ``Generates an original ET", the OSN ``Decides to notify", which contributes positively to the ``Generate more content".  The User B observes the proposed ET, and ``Decides to reply", which also contributes positively to the softgoal ``Generate more content".  The mechanism goes on.  If the OSN ``Decides to notify" the reply to other users, it contributes further to the softgoal ``Generate more content".

\section{Discussion} \label{sec:Chap_IStar_Extension_discussion}

\subsection{Benefits and Limitations} \label{sec:Chap_IStar_Extension_benefits_lim}

The motivating problem of this paper was the modelling of requirements for content recommendation on OSNs.  More specifically, we aimed at modelling the mechanism represented in Figure \ref{fig:Chap_IStar_Extension_cycle}.  We noticed that the original i-star did not allow us to model the dynamics observed on OSNs.  We also know that Petri Nets are a nice way to simulate the dynamic behavior of a system \cite{murata1989petri}.  We combined these two standards, using a layer mechanism to model, in one diagram, the requirements of a content RS. 
 
The benefits of our approach are threefold. 
 
Firstly, we do not introduce another extension, any new concepts, to an existing language.  Hence, the use of our proposal does not require any new learning.   

Secondly, the layer mechanism allows us to manage the complexity.  Each layer consists of a diagram using only the concepts and symbols of one language.  So, each layer can be read easily, but models the relevant information.  The combination of these two layers conveys the necessary information for the modelling of requirements for content recommendations. 
 
Thirdly, the nature of our approach, that is the use of layers, allows us to extend the scope of the models without any difficulty.  On the one hand, other languages can be used together via layers.  On the other hand, we can imagine adding more details to  our models by adding other layers, in order to manage more aspects of a RS for OSNs. 
 
The main limitations of our models are related to the dynamics.  Firstly, our diagrams show ``one instance" of the mechanism.  However, the fact that we can observe n instances of the mechanism could be modeled more clearly.  Secondly, we show the interaction between two users.  However, on OSNs many users are involved in many different mechanisms.  This situation does not show on our models here.

\subsection{Reasoning} \label{sec:Chap_IStar_Extension_reasoning}

We will now turn to the analysis of the model, using both forward and backward reasoning, that is, we apply a ``what if?" analysis.  We apply the procedure proposed by Horkoff \& Yu \cite{horkoff2009qualitative,horkoff2009evaluating} and by Giorgini et al. \cite{giorgini2002reasoning,giorgini2003formal,giorgini2005goal} to our model.  To nodes in the graph, we associate one of the following labels: Satisfied (S), Partially Satisfied (PS), Conflict (C), Unknown (U), Partially Denied (PD), and Denied (D). Sat(G) and Den(G) can range in: Full (F), Partial (P), and None (N) \cite{giorgini2005goal} .  
 
We consider three questions:
\begin{itemize}
\item What is the effect of the RS's decision to not notify the Receiver? 
\item What is the effect of the Receiver's decision to reply to the ET? 
\item What is the effect of the Receiver's decision to not reply to the ET? 
\end{itemize}
The results of the forward analysis are presented in Tables \ref{tab:Chap_IStar_Extension_reasoning1} to \ref{tab:Chap_IStar_Extension_reasoning3}.  We applied the propagation axioms proposed in \cite{horkoff2009qualitative,horkoff2009evaluating}.

\begin{table}
\begin{center}
 \caption{What is the effect of the RS's decision to not notify the Receiver?: Reasoning}
 \label{tab:Chap_IStar_Extension_reasoning1}
 \makebox[\textwidth][c]{
  \begin{tabular}{l||c|c||l||c|c}
 \hline
\textbf{Nodes} & \textbf{Initial}  & \textbf{Final}  & \textbf{Nodes}  & \textbf{Initial}  & \textbf{Final} \\
 \hline 
 Maintain desire to use OSN  & & PS (HJ) & Decision heuristics    & S &    \\  
 Allow users to generate ETI's  & & S & Decide to notify   & S &   \\ 
 Allow users to post Profile ETs & S &    & Decide to not notify   & D &    \\  
 Propose relevant recommendations & & S & Notifications are relevant  & & S   \\  
\cline{4-6}
Gather user information & & S  & Feature & & S\\ 
Analyze relations between users   & S &  & Original ETI  & & S  \\ 
Analyze content of ETI   & S & & Reply ETI  & & D \\ 
Gather ETI  & & S & Notification  & D S  \\ 
Gather original ETI & S &  & Use OSN (GD)  & & S  \\ 
Gather reply ETI & D &   & Use OSN (Task)   & S &   \\ 
Mitigate information overload  & & PS (HJ) & Generate ETI  & & S  \\ 
Improve user experience   & & S  & Generate original ETI   & S &   \\ 
Generate more content & & PS (HJ) & Generate reply ETI & D &  \\ 
\cline{4-6}
Encourage dynamics  & & D (HJ)  & Use OSN passively & & D  \\ 
Refine user profile  & & S  & Receive notification of user's ETI & & D  \\ 
\cline{1-3}
User information  & & S & Evaluate ETI  & & U  \\ 
Propose relevant notifications & & S & Evaluate ETI as relevant  & & U   \\ 
\cline{1-3}
Choose users to receive notifications  & & S & React to user's ETI  & & U   \\ 
Get user information  & & S & Evaluate ETI as irrelevant  & & U   \\ 
Apply decision heuristics & & S & Do not react to user's ETI  & & U   \\ 
\hline
  \end{tabular}}
\end{center}
\end{table}

\begin{table}
\begin{center}
 \caption{What is the effect of the Receiver's decision to reply to the ET?: Reasoning}
 \label{tab:Chap_IStar_Extension_reasoning2}
 \makebox[\textwidth][c]{
  \begin{tabular}{l||c|c||l||c|c}
 \hline
\textbf{Nodes} & \textbf{Initial}  & \textbf{Final}  & \textbf{Nodes}  & \textbf{Initial}  & \textbf{Final} \\
 \hline 
 Maintain desire to use OSN  & & PS (HJ) & Decision heuristics    & S &    \\  
 Allow users to generate ETI's  & & S  & Decide to notify   & S &   \\ 
 Allow users to post Profile ETs & S &    & Decide to not notify   & D &    \\  
 Propose relevant recommendations & & S & Notifications are relevant  & & S   \\  
\cline{4-6}
Gather user information & & S  & Feature & & S\\ 
Analyze relations between users   & S &  & Original ETI  & & S  \\ 
Analyze content of ETI   & S & & Reply ETI  & & S \\ 
Gather ETI  & & S & Notification  & & S  \\ 
Gather original ETI & S &  & Use OSN (GD)  & & S  \\ 
Gather reply ETI &  & S  & Use OSN (Task)   & S &   \\ 
Mitigate information overload  & & D & Generate ETI  & & S  \\ 
Improve user experience   & & S  & Generate original ETI   & S &   \\ 
Generate more content & & S & Generate reply ETI & & S \\ 
\cline{4-6}
Encourage dynamics  & & S (HJ)  & Use OSN passively & & S  \\ 
Refine user profile  & & S  & Receive notification of user's ETI & & S  \\ 
\cline{1-3}
User information  & & S & Evaluate ETI  & & S  \\ 
Propose relevant notifications & & S & Evaluate ETI as relevant  & & S   \\ 
\cline{1-3}
Choose users to receive notifications  & & S & React to user's ETI  & S & \\ 
Get user information  & & S & Evaluate ETI as irrelevant  & & D  \\ 
Apply decision heuristics & & S & Do not react to user's ETI  & D &   \\  
\hline
  \end{tabular}}
\end{center}
\end{table}

\begin{table}
\begin{center}
 \caption{What is the effect of the Receiver's decision to not reply to the ET?: Reasoning}
 \label{tab:Chap_IStar_Extension_reasoning3}
 \makebox[\textwidth][c]{
  \begin{tabular}{l||c|c||l||c|c}
 \hline
\textbf{Nodes} & \textbf{Initial}  & \textbf{Final}  & \textbf{Nodes}  & \textbf{Initial}  & \textbf{Final} \\
 \hline 
 Maintain desire to use OSN  & & PS (HJ) & Decision heuristics    & S &    \\  
 Allow users to generate ETI's  & & S  & Decide to notify   & S &   \\ 
 Allow users to post Profile ETs & S &    & Decide to not notify   & D &    \\  
 Propose relevant recommendations & & S & Notifications are relevant  & & S   \\  
\cline{4-6}
Gather user information & & S  & Feature & & S\\ 
Analyze relations between users   & S &  & Original ETI  & & S  \\ 
Analyze content of ETI   & S & & Reply ETI  & & D \\ 
Gather ETI  & & S & Notification  & & S  \\ 
Gather original ETI & S &  & Use OSN (GD)  & & S  \\ 
Gather reply ETI &  & D  & Use OSN (Task)   & S &   \\ 
Mitigate information overload  & & D & Generate ETI  & & S  \\ 
Improve user experience   & & S  & Generate original ETI   & S &   \\ 
Generate more content & & S & Generate reply ETI & & D \\ 
\cline{4-6}
Encourage dynamics  & & PD (HJ)  & Use OSN passively & & S  \\ 
Refine user profile  & & S  & Receive notification of user's ETI & & S  \\ 
\cline{1-3}
User information  & & S & Evaluate ETI  & & S  \\ 
Propose relevant notifications & & S & Evaluate ETI as relevant  & & D   \\ 
\cline{1-3}
Choose users to receive notifications  & & S & React to user's ETI  & D & \\ 
Get user information  & & S & Evaluate ETI as irrelevant  & & S  \\ 
Apply decision heuristics & & S & Do not react to user's ETI  & S &   \\  
\hline
  \end{tabular}}
\end{center}
\end{table}

The results of the forward reasoning show that the goal ``Maintain desire to use OSN" is, in every scenario, partially satisfied.  Given that multiple links contributed to the goal, human judgment (HJ) was necessary to label the goal \cite{horkoff2009qualitative,horkoff2009evaluating}.  The difference between the three scenarios shows in the final label of the following softgoals: Mitigate information overload, Generate more content, and Encourage dynamics. 
 
However, the analysis regarding the softgoals is even clearer when looking at the second layer.  The Petri Net shows the feedback loop, when the receiver decides to reply to the ET.  Indeed, the RS then applies the heuristic to the latter, and the dynamic continues.

\section{Related Work} \label{sec:Chap_IStar_Extension_rel_work}
OSNs have been largely explored in the literature. The contributions focus on the privacy and trust issues \cite{dwyer2007trust,strater2008strategies,guha2008noyb,golbeck2009trust,madejski2011failure}, the topological characteristics of social networks induced by the links between the members \cite{mislove2007measurement,ahn2007analysis,kumar2010structure,kwak2010twitter,xiang2010modeling,leskovec2010predicting,heer2005vizster,viswanath2009evolution,fu2008empirical,mislove2008growth}, the activities carried out by users on OSN \cite{guo2009analyzing,schneider2009understanding,benevenuto2009characterizing}, the user profile \cite{guo2009analyzing,schneider2009understanding,benevenuto2009characterizing,nosko2010all,emanuel2013does,silfverberg2011ll,alim2011online,kontaxis2011detecting,narendula2011my3,heckmann2005gumo,berkovsky2008mediation,abel2011analyzing,kapsammer2012user}, about the reasons why people become members of an OSN \cite{subrahmanyam2008online,livingstone2008taking,park2009being,raacke2008myspace,java2007we}, and the acceptance of OSN \cite{hsu2004people,hsu2008acceptance,gangadharbatla2008facebook}.  
 
However, there is little research of OSNs from the perspective of Requirements Engineering (RE). In previous work, we proposed RE patterns for the modeling of OSN features \cite{bouraga2014requirements}; and we also conducted an empirical study of features importance of OSNs. We also proposed an empirical study of notifications' importance for OSN users \cite{bouraga2015empirical}. We believe that this study gives us insight into the event types users want to see in priority; and this is relevant because the design of new systems, and OSNs in particular, involves deciding what to show to users. Our previous works differ from our approach here.  Because, firstly, we modeled existing features of OSNs \cite{bouraga2014requirements}, not taking into account any dynamics.  Secondly, we evaluated the content that a RS should consider in priority, while our work here focus on the requirements that the RS should satisfy.  Other studies related to the requirements for OSNs include Gurses et al. \cite{gurses2008privacy}, who addressed a specific requirements definition problem, namely the privacy requirements for OSNs. Their study results in an analytical framework for characterizing privacy breaches on OSNs.  The latter study is different from ours, because we do not focus solely on the privacy requirements, and on a privacy requirements ontology, but rather on various kinds of requirements an OSN should satisfy.
 
Finally, there is considerable work on extending requirements modeling languages.

Souza et al. \cite{souza2012requirement,souza2012requirements} proposed the AwReqs and EvoReqs. They applied to a goal model loosely based on an i-star model.  AwReqs are requirements that talk about the states assumed by other requirements at runtime.  They represent undesirable situations to which stakeholders would like the system to adapt in case they happen.  They can be used as indicators of requirements convergence at runtime.  AwReqs indicate the situation that require adaptation. EvoReqs prescribe what to do in these situations.  They consist of specific changes to be carried out on the requirements model, under specific circumstances.  They are modeled as Event-Condition-Action (ECA) rules that are activated if an event occurs and a certain condition holds.

The AwReqs would be: 
\begin{itemize}
\item Notifications are relevant: Never Fails
\item Mitigate information overload: Maximum notification rate per hour
\item Generate more content: Minimum notification rate per hour
\item Encourage dynamics: Never fails
\item Refine user profile: Never fails
\end{itemize}

While the EvoReqs (ER) would be: 
\begin{enumerate}
\item ER1: 
	\begin{enumerate}
	\item Event: Send notification
	\item Condition: Decide to reply
	\item Actions
			\begin{enumerate}
			\item Roles switch
			\item Generate reply
			\item Gather reply
			\end{enumerate}
	\end{enumerate}
\item ER2: 
	\begin{enumerate}
	\item Event: Minimum rate of notifications $<$ Threshold
	\item Condition: new ETI flagged as irrelevant
	\item Actions
			\begin{enumerate}
			\item Notify irrelevant content
			\item Suspend(Notifications are relevant)
			\end{enumerate}
	\end{enumerate}
\item ER3: 
	\begin{enumerate}
	\item Event: Maximum rate of notifications $>$ Threshold
	\item Condition: a.	new ETI flagged as relevant
	\item Action: Not notify relevant content
	\end{enumerate}
\end{enumerate}

Ali et al. \cite{ali2013reasoning} addressed the reasoning with contextual requirements.  The authors define context as ``a partial state of the world that is relevant to an actor's goals".  In our example, the contexts would be: 
\begin{enumerate}
\item The decision heuristic evaluates the ET as irrelevant 
\item The decision heuristic evaluates the ET as relevant
	\begin{enumerate}
	\item Decide to notify (RS), AND
		\begin{enumerate}
		\item Apply decision heuristic, AND
		\item Evaluation of the relevance = Core
		\end{enumerate}
	\item Receive notification (User)
	\end{enumerate}
\item The receiver decides to reply to the original ET
	\begin{enumerate}
	\item React to the user's ET, AND
		\begin{enumerate}
		\item Evaluate the ET as relevant, AND
		\item Decide to reply
		\end{enumerate}
	\item Generate reply, AND
	\item Gather reply
	\end{enumerate}
\end{enumerate}

These extensions offer nice solutions, however, we believe that the results is not as comprehensive or as appropriate for our problem as our proposal in this paper.  The connection between the various concepts, that is, the AwReqs and EvoReqs for the proposal of Souza et al. \cite{souza2012requirement,souza2012requirements}, and the contexts for Ali et al. \cite{ali2013reasoning} is not as integrated as ours.  For the former solution, Souza et al. (\cite{souza2012requirement,souza2012requirements} model the AwReqs on the requirements model, but the EvoReqs are rules specified in natural language outside of the diagram.  For the latter solution, Ali et al. \cite{ali2013reasoning} clearly manage the context, and integrate this notion directly in the models.  However, the notion of dynamics is not included, whereas we propose an integrated solution, including the dynamics aspect.  
 
Other examples of extensions include the following.  Liaskos et al. \cite{liaskos2010integrating} proposed an extension of traditional goal modelling notation in order to be able to model the distinction between mandatory and optional goals. Van Lamsweerde \& Letier \cite{van1998integrating} proposed to integrate techniques for reasoning about obstacles to the satisfaction of goals, requirements, and assumptions, into KAOS, a methodology developed for the support of the whole process of requirements elaboration. Mouratidis et al. \cite{mouratidis2002using,gani2003analysing,mouratidis2003ontology,mouratidis2006secure,mouratidis2007secure,mouratidis2009enhancing,mouratidis2011secure} propose an extension of the Tropos Methodology in order to integrate security concerns in agent-oriented methodologies. Qureshi et al. \cite{qureshi2011requirements} addressed RE for self-adaptive systems. Using requirements-aware systems, and taking advantage of the Techne modeling language, they proposed a framework to ``enable continuous adaptive requirements engineering (CARE) for SAS". Closer to our work, Alencar et al. \cite{alencar2006identifying} (2006a, 2006b) discovered that i-star did not allow to explicitly and systematically handle the crosscutting nature of some concerns in agent-oriented modelling. In order to bridge this gap, they aim at the extension of the i-star framework.  Liaskos \& Mylopoulos \cite{liaskos2010temporally} added temporal annotations to goal models, and similarly Wang \& Lesp\'erance \cite{wang2001agent} added process specification annotations to SR diagrams of i*.  These extensions are interesting and fall into our problem domain, but we do not want annotations to the original i* model, we believe that a layer mechanism is more convenient to manage complexity. 
 
Considerable work also exists regarding transformation/mapping of goal models.  Horkoff et al. \cite{horkoff2015using} conducted a literature review of goal-oriented techniques that introduce a transformation/mapping/integration to a software artifact, and identified 49 papers addressing i* transformation/mapping.  A relevant example of such a mapping/transformation was proposed by Grau et al. \cite{grau2008prim}.  The authors conduct process reengineering based on i*: PRiM.  They start from a process description, and therefrom derive an i* model.  This is relevant, because they have to translate process constructs in their i* models; and here we are trying to translate some dynamic in ours.  However, the motivation behind PRiM is different from ours.  With PRiM the authors aim for the specification of the new system, that is, they follow a set of guidelines for constructing Use Case diagram from the i* model.  Our scope here does not cover as much as PRiM, here we are only concerned with the representation of the problem domain.

\section{Conclusion and Future Work} \label{sec:Chap_IStar_Extension_ccl}

We believe i-star is appropriate for the modeling of OSN requirements. However, as mentioned above, the existing concepts in i-star do not allow us to model the dynamics observe in the use of OSNs. Hence, we proposed an ``add-on" to the existing framework, by introducing a second layer. The latter consists of a Petri Net modelling the dynamics observed in OSNs. 
 
In this paper, we started by presenting the case study and our approach of using layers.  We then described the languages we considered, followed by the corresponding models.  We then apply our approach.  We discussed the results and the practical implications of our work by showing that the proposed contribution is not limited to OSN, but can also be extended to other systems that share similar features; such as ESM. 
 
Future work will consist in addressing the limitations we raised in Section \ref{sec:Chap_IStar_Extension_benefits_lim}.  More specifically, we aim at providing a more general model, taking into account the various mechanisms an individual user can be involved in; as well as the several instances of mechanisms that can exist.

\bibliographystyle{splncs} 
\bibliography{bibli_i_star_extension}

\end{document}